\documentclass{llncs}
\usepackage{makeidx}  
\usepackage{psfig}
\begin{document}
\frontmatter          
\pagestyle{headings}  
\addtocmark{Hamiltonian Mechanics} 
\mainmatter              
\title{Lyapunov instability and collective tangent space
dynamics of fluids}
\titlerunning{Lyapunov instability}  
%
\author{Harald A. Posch  \and Christina Forster}
\authorrunning{H. A. Posch et al.}   
\tocauthor{ Harald A. Posch, Christina Forster ( University of Vienna )}
\institute{Institut f\"ur Experimentalphysik, Universit\"at Wien, \\
Boltzmanngasse 5, A-1090 Vienna, Austria}

\maketitle              

\begin{abstract}
The phase space trajectories of many body systems charateristic of simple 
fluids are highly unstable. We quantify this instability  
by a set of Lyapunov exponents, which are the rates of exponential divergence, 
or convergence, of initial (infinitesimal) perturbations along 
carefully selected 
directions in phase space.  It is demonstrated that the perturbation 
associated with the {\em maximum} Lyapunov exponent is localized
in space. This localization persists in the large-particle
limit, regardless of the interaction potential. The perturbations belonging
to the {\it smallest positive} exponents, however, are sensitive to the
potential. For hard particles they form well-defined long-wavelength modes.
The modes could not be observed for systems interacting with a soft
potential due to surprisingly large fluctuations of the local 
time-dependent exponents.
\end{abstract}
\section{Lyapunov spectra}
Recently, molecular dynamics simulations have been used to study many 
body systems representing simple fluids or solids from the point of view of
dynamical systems theory. Due to the convex dispersive surface of the atoms, 
the phase trajectory of such systems is highly unstable and 
leads to an exponential growth, or decay, of small (infinitesimal) 
perturbations of an initial state along specified directions in phase space. 
This so-called Lyapunov instability is described by a set of rate constants, 
the Lyapunov exponents $\{\lambda_l,  l=1,\ldots,D\} $, to which we
refer as the Lyapunov spectrum. Conventionally, the exponents are 
taken to be ordered by size, $\lambda_l \ge \lambda_{l+1}$.  There are 
altogether $D = 2dN$ exponents, where $d$ and $D$ denote the dimensions of 
space and of phase space, respectively, and $N$ is the number of particles.
For fluids in nonequilibrium 
steady states  close links between the Lyapunov spectrum and  macroscopic 
dynamical properties, such as transport coefficients, irreversible
entropy production, and the Second Law of thermodynamics, have been 
established \cite{ph88,PH89,em90,HOO,GASP,DO99}. This important result
provided the motivation for us
to examine the spatial structure of the perturbed states associated 
with the various exponents. Here we present some of our results for 
two simple many-body systems representing dense two-dimensional fluids in
thermodynamic equilibrium. The first model consists of $N$ hard disks (HD)
interacting with hard elastic collisions, the second of $N$ soft
disks interacting with a purely repulsive Weeks-Chandler-Anderson (WCA) 
potential.

The instantaneous  state of a planar particle system is given by the
$4N$-dimensional phase space vector  $ {\bf \Gamma} = \{{\bf r}_i, {\bf p}_i,
; i = 1,\ldots,N \}$, where ${\bf r}_i$ and ${\bf p}_i$ denote the
respective position and linear momentum of molecule $i$. An infinitesimal
perturbation $ \delta {\bf \Gamma} = \{\delta{\bf r}_i, \delta {\bf p}_i;
i = 1,\dots,N \}$ evolves according to motion equations obtained by
linearizing the equations of motion for ${\bf \Gamma}(t)$.
For ergodic systems there exist $D = 4N$ orthonormal 
initial vectors  $ \{ \delta {\bf \Gamma}_l(0); l = 1,\ldots ,4N\}$
in tangent space, such that the Lyapunov exponents
\begin{equation}
  \lambda_l = \lim_{t\to\infty}\frac{1}{t} \ln
              \frac{|\delta {\bf\Gamma}_l(t)|}{|\delta {\bf\Gamma}_l(0)|}
              \;\; , \;\;l = 1,\ldots ,4N   .
\label{lyap}
\end{equation}
exist and are independent of the initial state 
\cite{oseledec,eckmann}. Geometrically,
the Lyapunov spectrum describes the stretching and contraction along 
linearly-independent phase space directions of an infinitesimal hypersphere 
co-moving with the flow.
For equilibrium systems the symplectic nature of the motion equations
assures the conjugate pairing rule to hold \cite{Ruelle}: 
the exponents appear in pairs with a vanishing pair sum,
 $\lambda_{l} + \lambda_{4N+1 - l} = 0$.
Thus, only the positive half of the spectrum $\{\lambda_{1 \le l \le 2N}\}$
needs to be calculated.  The sum 
of all Lyapunov exponents vanishes, which according to Liouville's theorem
is a manifestation of the conservation of phase volume by Hamiltonian
systems.  Six of the exponents,  
$\{\lambda_{2N-2 \le l \le 2N+3}\}$, always vanish as a consequence
of the conservation of energy, momentum, and center of mass, and of
the non-exponential time evolution of a perturbation vector parallel to the
phase flow. 

For the computation of a complete spectrum a variant of the classical 
algorithm by Benettin {\em et al.}  and Shimada {\em et al.}
is commonly used \cite{Benettin,Shimada}. It requires following
the time evolution of the 
reference trajectory and of an orthonormal set of tangent vectors  
$ \{\delta {\bf \Gamma}_l(t); l = 1,\ldots,4N \}$, where the latter
is periodically re-orthonormalized with a Gram-Schmidt (GS) procedure after 
consecutive  time intervals $\Delta t_{GS}$.
The Lyapunov exponents are determined from the time-averaged 
renormalization factors.  For the hard disk systems the free evolution 
of the particles is interrupted by hard elastic collisions and a linearized
collision map needs to be calculated. An algorithm based on this idea has
been developed by Dellago {\em et al.} \cite{DPH96}. Although we make use of 
the conjugate pairing symmetry and compute only the positive branch of the 
spectrum, we are presently restricted to about 1000 particles  by our available
computer resources.

   For our numerical work reduced units are used. In the case of the
Weeks-Chandler-Anderson interaction potential,
\begin{equation}
\phi(r)=\left\{
    \begin{array}{ll}
        4\epsilon[(\sigma/r)^{12}-(\sigma/r)^{6}]+\epsilon, & r<2^{1/6}\sigma\\
        0, & r\ge 2^{1/6}\sigma.
    \end{array}\right. ,
\end{equation}
the particle mass $m$, the particle diameter $\sigma$, and the time 
$(m \sigma^2/\epsilon)^{1/2}$ are unity. In this paper we restrict our 
discussion to a thermodynamic state with a total energy per particle, 
$E/N$, also 
equal unity.  For the hard-disk fluid, $ (Nm \sigma^2/K)^{1/2}$ is 
taken as the unit of time, where $K$ is the total 
kinetic energy, equal to the total energy $E$
of the system. There is no potential energy in this case.
For the reduced temperature we have $T = K/N$, where Boltzmann's constant 
is also unity.  In the following, Lyapunov exponents for the two model 
fluids will be compared for equal temperatures (and not for equal total 
energy).  This requires a rescaling of the hard-disk exponents by a factor of 
$\sqrt{(K/N)_{WCA}/(K/N)_{HD}}$ to account for
the difference in temperature. All our simulations are for a reduced
density $\rho \equiv N/V = 0.7$, where the simulation box is a square
with a  volume $V = L^2$ and a side length $L$. Periodic boundaries are used 
throughout.
\begin{figure}
\centerline{\psfig{angle=-90,width=12cm,file=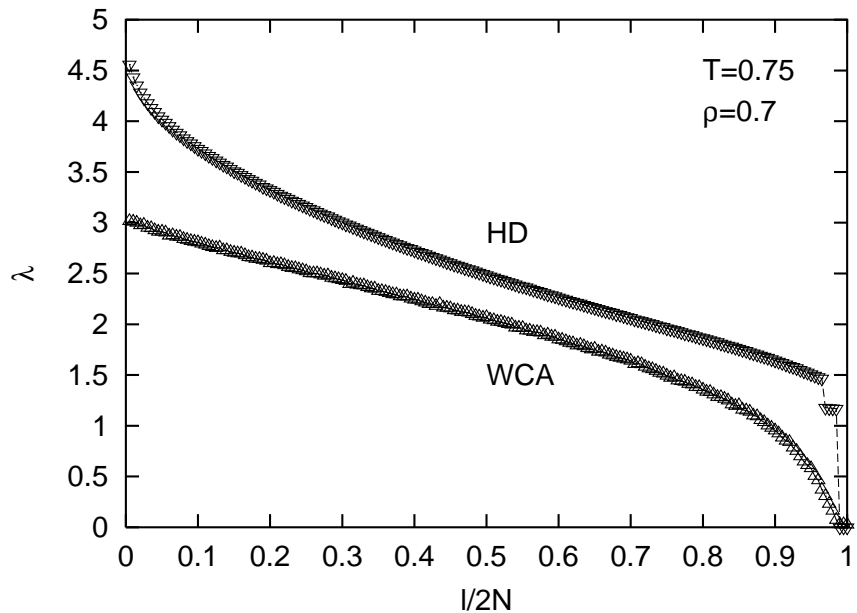}}
\caption{Lyapunov spectrum of a dense two-dimensional fluid consisting
of $N=100$ particles at a density $\rho = 0.7$ and a temperature
$T = 0.75$. The label WCA refers to a smooth Weeks-Chandler-Anderson
interaction potential, and HD indicates the hard disk system.}
\label{Fig1}
\end{figure}

    As an example, in Fig. \ref{Fig1} we compare the Lyapunov spectrum of a  
100-particle WCA fluid to a spectrum for an analogous hard disk system at 
the same temperature ($T = 0.75$) and density ($\rho =0.7$).  A reduced
index $l/2N$ is used on the abscissa.  It is surprising that the two spectra 
differ so much in shape and size. The difference persists in the 
thermodynamic limit.  The step-like structure displayed by the hard disk 
spectrum for $l/2N$ close to 1 is an indication of a coherent wave-like
shape of the associated perturbations, to which we refer as Lyapunov modes
\cite{ph00,MP02}.  We defer the discussion of these modes to Section 4.  
\section{Time-dependent local exponents}
We infer from Equ. (\ref{lyap}) that the Lyapunov exponents 
are time averages over an (infinitely) long
trajectory and, therefore,  are global properties of the system. They
can be rewritten as \cite{ph88,hdp02}
\begin{equation}
\lambda_l = \lim_{ \tau \to \infty}\int_{0}^{\tau} {\lambda'}_l
({\bf \Gamma}(t)) dt \equiv \langle  {\lambda'}_l \rangle,
\end{equation}
where the (implicitly) time-dependent function 
$ {\lambda'}_l({\bf \Gamma})$ depends on the state  
${\bf \Gamma}(t)$ which the system occupies in phase space 
at time $t$. Thus, $ {\lambda'}_l({\bf \Gamma})$ is called a local
Lyapunov exponent. It may be estimated from
\begin{equation}
{\lambda'}_l({\bf \Gamma}(t)) =\frac{1}{\Delta t_{GS}}
 \ln \frac{|\delta{\bf \Gamma}_l({\bf \Gamma}(t+ \Delta t_{GS})|}
          {|\delta{\bf \Gamma}_l({\bf \Gamma}(t)|},
\label{local}
\end{equation}
where $t$ and $t+ \Delta t_{GS}$ refer to times immediately
after consecutive Gram-Schmidt re-orthonormalization steps.
Its time average along a trajectory is denoted by $\langle \cdots \rangle$
and gives the global exponent $\lambda_l$.

\begin{figure}
\centerline{\psfig{angle=-90,width=12cm,file=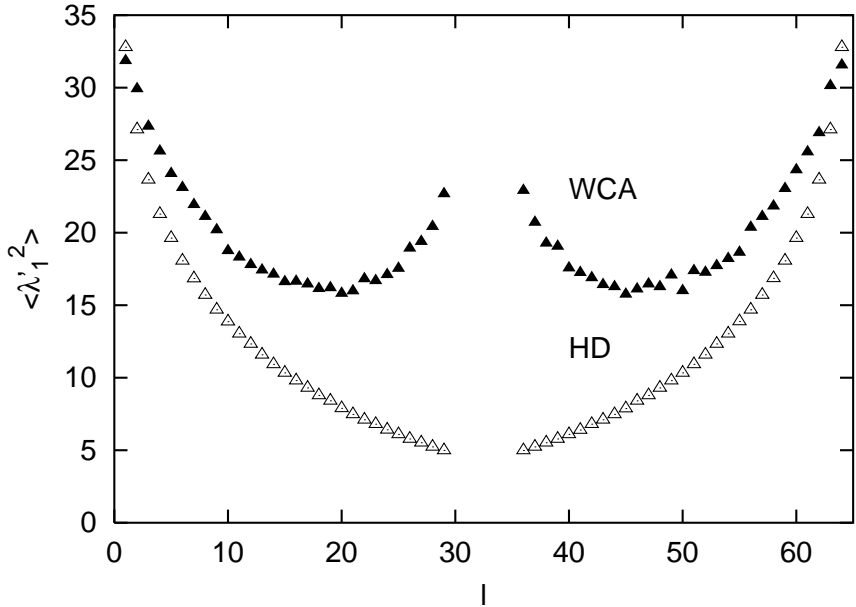}}
\caption{Second moment of all local Lyapunov exponents for
a planar 16-particle system. The labels WCA and HD refer to the
Weeks-Chandler-Anderson and hard disk models, respectively. 
The re-orthonormalization time $\Delta t_{GS} = 0.075$, 
the temperature $T= 0.75$, and the density $ \rho = 0.7$.}
\label{Fig2}
\end{figure}

The local exponents fluctuate considerably along a trajectory. This is 
demonstrated in Fig. \ref{Fig2}, where we have plotted 
the second momens $\langle  {\lambda'}_l^2 \rangle$ as a function of the
Lyapunov index $l, 1 \le l \le 4N$,  for a system of 
16 particles, both for the WCA and HD models. $l=1$ refers
to the maximum, and $l = 64$ to the most-negative exponent.
The points for $30 \le l \le 35$ correspond to the 6 vanishing exponents 
and are not shown. We infer from this figure that for the soft WCA particles
the fluctuations of the local exponents become very large
for the Lyapunov exponents describing relatively-weak instabilities with
near-zero growth rates, $l \to 2N$. For the hard disk system, however, 
the relative importance of the fluctuations becomes minimal for the same 
exponents.  We shall return to this point in Section 4.

\begin{figure}
\centerline{\psfig{angle=-90,width=10cm,file=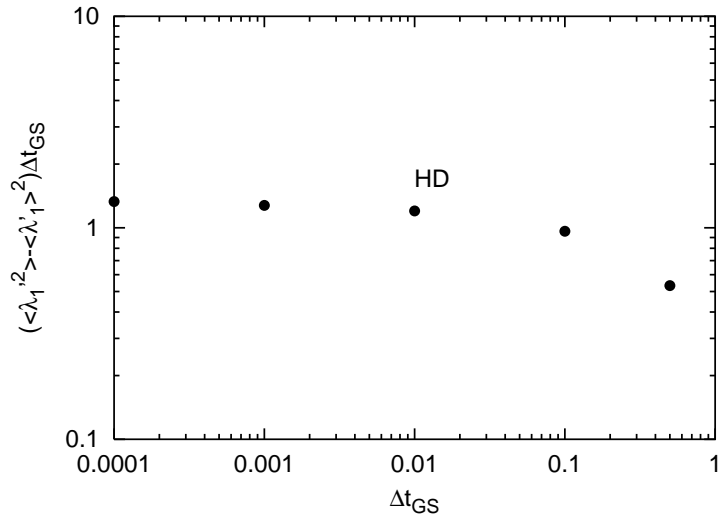}}
\caption{ Mean-square fluctuations of the local exponent ${\lambda'}_1$,
$\langle  {\lambda'}_1^2 \rangle - \langle  {\lambda'}_1 \rangle^2$,
times  $\Delta t_{GS}$,  as a function of
the re-orthonormalization interval $\Delta t_{GS}$. The system consists
of 64 hard disks.  The  temperature $T = K/64 = 1$, and the density
$\rho = 0.7$. }
\label{Fig3}
\end{figure}

   We note that the computation of the second moments  
$\langle  {\lambda'}_l^2 \rangle$ for the hard disk system requires 
some care.  Due to the hard core collisions they depend strongly on 
$\Delta t_{GS}$.  The variance of the fluctuating local exponents, 
$\langle  {\lambda'}_l^2 \rangle - \langle  {\lambda'}_l \rangle^2$,
varies with $1/\Delta t_{GS}$ for small $\Delta t_{GS}$,
as is demonstrated in Fig. \ref{Fig3} for $l = 1$. 
However, the shape of the fluctuation spectrum is hardly affected by the
size of the renormalization interval.

\section{The maximum exponent}
The maximum Lyapunov exponent is the rate constant for the
fastest growth of a phase space perturbation in a system. Thus, it is
dominated by the fastest dynamical events in the fluid, binary collisions. 
There is strong numerical evidence for the existence of the 
thermodynamic limit $\{ N,V \to \infty, N/V$constant$\}$ for
$\lambda_1$ and, hence, for the whole spectrum \cite{DPH96,MP02}. 
Furthermore, the perturbation belonging to $\lambda_1$ is strongly localized 
in space \cite{HBP98,MPH98b,MP02}. 
This may be
demonstrated by projecting the tangent vector $\delta {\bf \Gamma}_1$
onto the four-dimensional subspaces spanned by the perturbation components 
of the individual particles.  The squared norm of this projection,
$ \delta_i^2(t) \equiv (\delta {\bf r}_i)_l^2 + (\delta {\bf p}_i)_l^2 $,
indicates how active a particle $i$ is engaged in the growth process
of the pertubation associated with $\lambda_1$ \cite{MPH98a,MPH98b}.

\begin{figure}
\centerline{\psfig{angle=-90,width=12cm,file=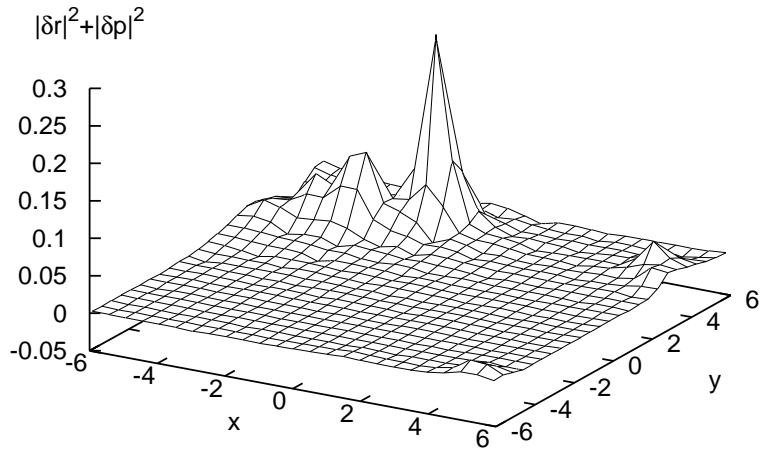}}
\caption{The surface depicts the instantaneous contributions of the
individual particles, plotted at the position of the particles in the 
simulation cell, to the squared norm of the perturbation vector
associated with the maximum exponent $\lambda_1$. The system consists
of 100 hard disks.
The temperature $T = K/N = 1$, and the density $\rho = 0.7$.}
\label{Fig4}
\end{figure}

In Fig. \ref{Fig4} $ \delta_i^2(t)$ 
is plotted along the vertical axis for all particles of a hard disk system 
at the respective positions $(x_i,y_i)$ of the disks in space, 
and the ensuing surface is interpolated over a periodic grid covering 
the simulation box. The figure refers to an instantaneous condition for
a well-relaxed system, and no averaging is involved. A strong localization 
of the active particles is observed at any instant of time.
Similar, albeit slightly broader peaks are observed for the WCA system
or other soft disk potentials.

   This localization is a consequence of two mechanisms: firstly,
after a collision the
delta-vector components of two colliding molecules
are linear functions of their pre-collision values and have only a
chance of further growth if their values before the collision
were already far above average. Secondly, each renormalization step
tends to reduce the (already small) components of the
other non-colliding particles even further. Thus, the competition for
maximum growth of tangent vector components favors the
collision pair with the largest components. The active zone moves in space
in a diffusive manner.

   The localization also persists in the thermodynamic limit. To
show this we follow Milanovi\'c {\em et al.} \cite{MP02} and order
all squared components $[\delta{\bf\Gamma}_1]^2_j; j = 1,\dots,4N$ 
of the perturbation vector $\delta {\bf \Gamma}_1$ according to size.
By adding them up, starting with the largest, we determine
the smallest number $A$ of terms,  required for the sum
to exceed a threshold $\Theta$. Then, $ C_{1,\Theta} \equiv A/4N$
may be taken as a relative measure
for the number of components actively contributing to $\lambda_1$:
\begin{equation}
        \Theta \le\left\langle \sum_{s=1}^{4N C_{1,\Theta}}
        \left[\delta{\bf\Gamma}_1\right]^2_s\right\rangle,
        \;\;\;\;\;\;\;\;
        [\delta{\bf\Gamma}_1]^2_i \ge [\delta{\bf\Gamma}_1]^2_j \;{\rm
        for}\; i<j.
\end{equation}
\begin{figure}
\centerline{\psfig{angle=-90,width=12cm,file=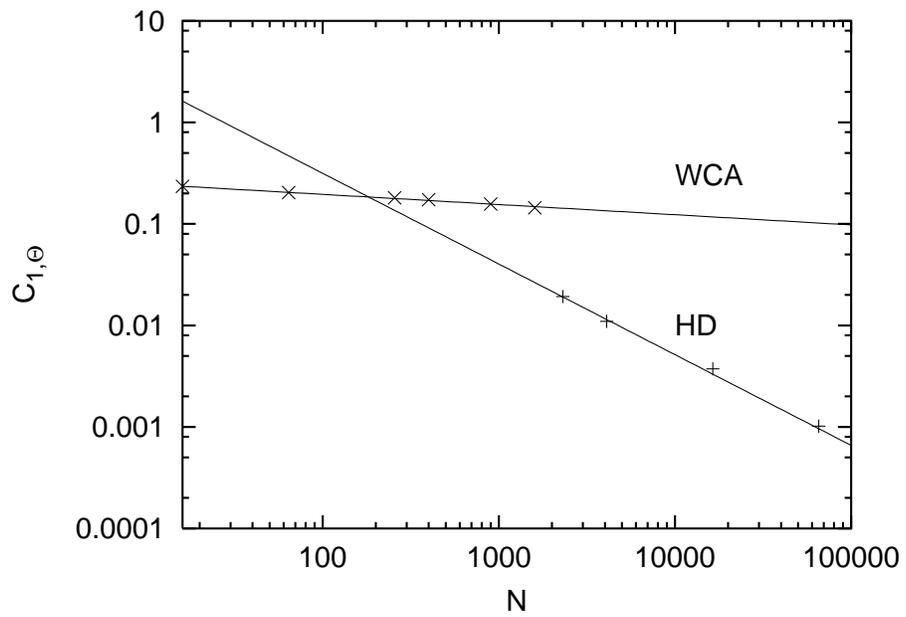}}
\caption{Dependence of the localization measure,
$C_{1,\Theta}$, on the number of particles $N$ for 
the parturbation associated with the maximum Lyapunov exponent $\lambda_1$.
The threshold $\Theta = 0.90$. The soft and hard disk results are labeled
by WCA and HD, respectively.  }
\label{Fig5}
\end{figure}
Obviously, $C_{1,1}=1$.

In Fig. \ref{Fig5} $C_{1,\Theta}$ is shown for
$\Theta = 0.90$ as a function of the particle number $N$, both for the WCA fluid
and for the hard disk system. It converges to zero if our data are 
extrapolated to the thermodynamic limit, $N\to\infty$.  
This supports our assertion that in an infinite system only a
{\em vanishing} part of the tangent-vector components (and, hence, 
of the particles)  contributes significantly to the
maximum Lyapunov exponent at any instant of time. 
\section{Lyapunov modes}
We have already mentioned the appearance of a step-like structure in the
Lyapunov sepctrum of the hard disk system for the positive exponents closest 
to zero, which was first observed in Ref. \cite{DPH96}.  
They are a consequence of coherent wave-like spatial patterns
generated by the  perturbation vector components associated with the 
individual particles \cite{ph00,MP02}.

\begin{figure}
\centerline{\psfig{angle=-90,width=12cm,file=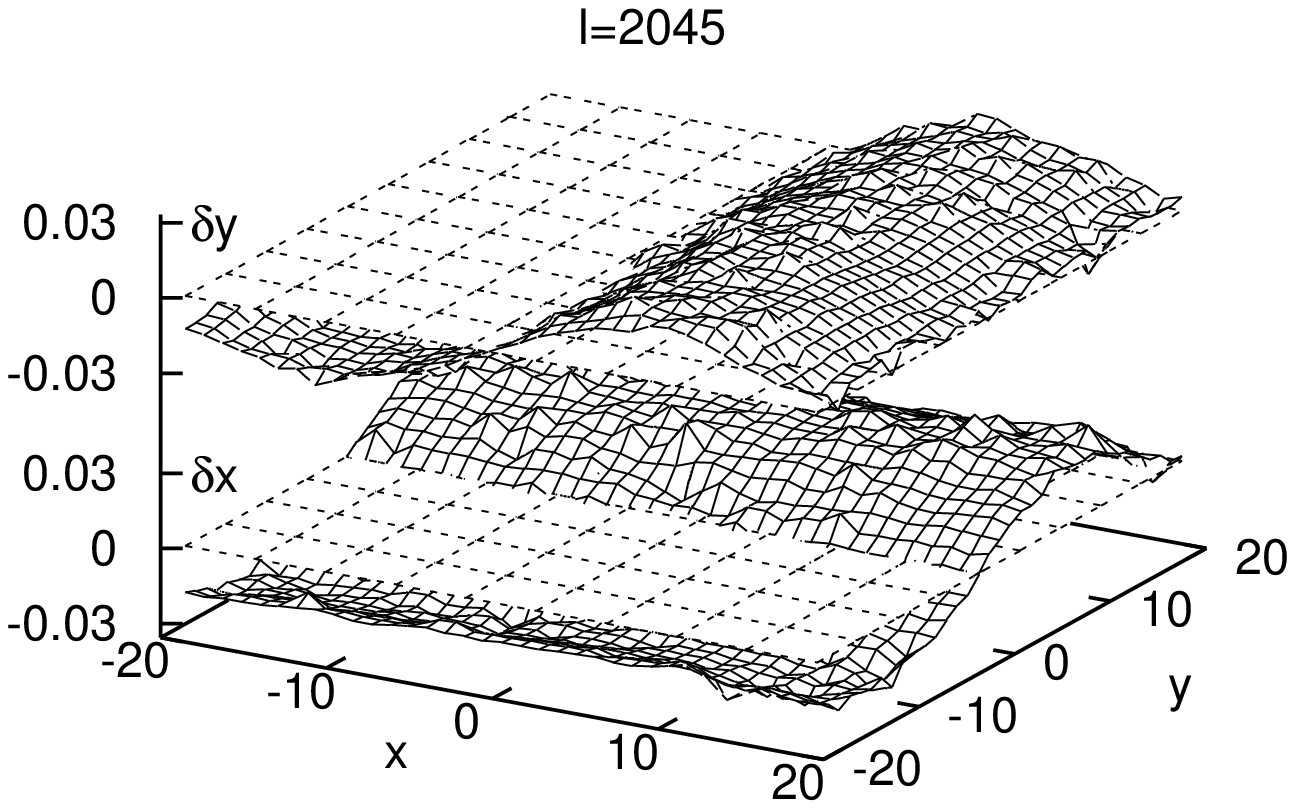}}
\caption{Transverse Lyapunov mode of type $T_1$ associated with
the smallest positive exponent $\lambda_{2045}$ for 1024 hard disks
at a density $\rho = 0.7$. The temperature $K/N = 1$. 
\newline
Bottom: Plot of the tangent vector components $\delta x$ of the
individual particles (along the vertical axis) for all particles
at their instantaneous postions $(x,y)$ in the simulation box
(horizontal plane).
Top: Analogous plot for $\delta y$.}
\label{Fig6}
\end{figure}

\noindent
In Fig. \ref{Fig6} this is visualized by plotting the
perturbations in the $x$ (bottom surface) and $y$ directions (top surface),
$\{\delta x_i, i=1,\dots,N\}$ and $\{\delta y_i, i=1,\dots,N\}$, 
respectively, along the vertical axis
at the instantaneous particle positions $(x_i,y_i)$ of all particles $i$. 
This figure depicts a  transverse Lyapunov mode of type $T_1$, for which 
the perturbation components are 
perpendicular to the respective wave vectors with one wave length
equal to $L$, the linear extension of the simulation box.
The system consists of $N = 1024$ hard disks, and the perturbation 
vector considered  in the figure belongs to the smallest positive 
exponent $\lambda_{2045}$.  An analogous plot of $\delta p_{x}$ and  
$\delta p_{y}$ for the same perturbation vector is identical in shape to
that of $\delta x$ and $\delta y$ in Fig. 6, with the same phases of the
waves. This is a consequence of the fact that the perturbations
are solutions of linear first-order  equations instead of second
order equations such as the wave equation. 
Furthermore, the exponents for $2042 \le l \le 2045$ are equal. This
four-fold degeneracy of non-propagating transversal modes, and an 
analogous eight-fold degeneracy of propagating longitudinal modes, 
are responsible for a complicated step structure for $l$ close to $2N$,
which has been studied in detail in Refs. \cite{ph00,fmp02}.  

The wave length of the modes and the value of the corresponding exponents
are determined by the size $L$ of the square simulation box. There 
is a kind of linear dispersion relation \cite{ph00,MP02} indicating that
the smallest positive exponent is proportional to $1/L$. This assures that
for a square simulation box with aspect ratio 1 there is no
positive lower bound for the positive exponents of a hard disk system
in the thermodynamic limit \cite{MP02,fmp02}. 

   So far, our discussion of modes applies only to hard disk fluids. In spite 
of a considerable computational effort we have not yet been able to indentify
modes for
two-dimensional fluid systems with a soft interaction potential
such as WCA or similar potentials \cite{hpfdz02}. 
This surprising result seems to be due to the very strong fluctuations 
of the local exponents discussed
in Section 2. They tend to obscure any mode in the system in spite 
of considerable averaging and make a positive identification very difficult.

  We are grateful to Christoph Dellago, Robin Hirschl, Bill Hoover, and
Ljubo\-mir Milanovi\'c for many illuminating discussions. This work was 
supported by the Austrian Fonds zur F\"orderung der wissenschaftlichen
Forschung, Grants P11428-PHY and P15348-PHY.

%

\end{document}